\documentstyle[12pt]{article}
\newcommand{\newc}{\newcommand}
\newcommand{\av}{\varepsilon}
\newcommand{\bv}{\overline{\varepsilon}}
\newc{\hmu}{\hat{\mu}}\newc{\hnu}{\hat{\nu}}
\newc{\hsg}{\hat{\sigma}}\newc{\hrho}{\hat{\rho}}
\newc{\htau}{\hat{\tau}}\newc{\hkp}{\hat{\kappa}}
\newc{\hio}{\hat{\iota}}\newc{\hep}{\hat{\epsilon}}
\newc{\hth}{\hat{\theta}}\newc{\hpi}{\hat{\pi}} \newc{\kp}{\kappa}

\newcommand{\al}{\alpha} \newcommand{\bt}{\beta}
\newcommand{\g}{\gamma} \newcommand{\dl}{\delta}
\newcommand{\e}{\epsilon} \newcommand{\lb}{\lambda}
 
\newcommand{\Sg}{\Sigma}\newcommand{\T}{{\cal T}}
\newcommand{\ha}{\hat{a}} 
\newc{\hc}{\hat{c}} \newcommand{\ba}{\begin{eqnarray}}
\newcommand{\ea}{\end{eqnarray}}
\newcommand{\be}{\begin{equation}}
\newcommand{\ee}{\end{equation}}
\newc{\dr}{{\rm d}}
\begin{document}

\begin{titlepage}
\begin{tabbing}
   \hspace{13.4cm} \=    \kill \>   \=  IUCAA 28/94 \+\+ \\ January 1995
 \\ gr-qc/9409022
\end{tabbing}\begin{center}
\Large{{\bf Spinors and the reference point}\footnote{to appear in
{\em Classical and Quantum Gravity}}\\
{\bf of quasilocal energy}}

\bigskip \bigskip \normalsize Stephen R. Lau\footnote{email:
lau@iucaa.ernet.in}

\bigskip {\it Inter-University Centre for\\
Astronomy and Astrophysics\\ Post Bag 4, Ganeshkhind\\
Pune, India 411 007}

\begin{abstract}
This paper investigates the relationship between the quasilocal energy of Brown
and York and certain spinorial expressions for gravitational energy constructed
from the Witten-Nester integral. A key feature of the Brown-York method for
defining quasilocal energy is that it allows for the freedom to assign the
reference point of the energy. When possible, it is perhaps most natural to
reference the energy against flat space, i.e. assign flat-space the zero
value of energy. It is demonstrated that the Witten-Nester integral when
evaluated
on solution spinors to the Sen-Witten equation (obeying appropriate
boundary conditions) is essentially the
Brown-York quasilocal energy with a reference point
determined by the Sen-Witten spinors. For the case of round spheres in the
Schwarzschild geometry, these spinors determine the flat-space reference point.
A similar viewpoint is proposed for the Schwarzschild-case quasilocal energy of
Dougan and Mason.
\end{abstract}
\end{center}
\end{titlepage}

\section{Introduction}
The ``local'' concept of energy in general relativity is quasilocal or,
in other words, associated with closed two-surfaces in spacetime.
Recently, many attempts to
define quasilocal energy with spinor methods have relied on the
so-called ``Witten-Nester two-form'' (famous from the Witten-style
proof \cite{Witten} of the positive energy theorem in the
asymptotically-flat scenario). When the two-form is integrated over
a generic two-surface $B$, the resulting integral may be viewed as a
functional of $B$ spinor fields. The goal is then to determine which
are the ``correct'' spinors to be inserted into this functional such
that it serves as the quasilocal energy. Further, the assumption is
that the correct spinors arise as solutions to some supplementary
equation (which may or may not be the Sen-Witten
equation\cite{Witten,Sen} used in the asymtotically-flat scenario).

Though this approach for defining quasilocal energy has been the
focus of considerable research, several aspects of such spinor
constructions have remained unclear. For instance, the approach
seems to make essential use of a spinor with one index. This
appears to imply that spinors are somehow essential for defining
quasilocal energy, in contrast to the asymptotically-flat scenario
where the ADM energy
\cite{ADM} may be defined without spinors. Another problem in the
quasilocal context is
the absence of a rigorous interpretation of the Witten-Nester
integral as a boundary value of the gravitational
Hamiltonian.\cite{FM}\footnote{Whether or not it is
possible to find a truly satisfactory Hamiltonian for a spatially
bounded slice $\Sigma$ is a subtle issue in its own right. Following
Ref. \cite{BY}, this paper assumes that the correct Hamiltonian for a
bounded region is the one which is ``read off'' from the canonical form of
the gravitational action appropriate for a spatially bounded spacetime
region.} Yet another issue is whether or not the supplementary spinor
equation should be independent of the spanning three-slice $\Sg$,
so that, as is often claimed, the quasilocal energy is independent
of the $\Sg$ chosen to span $B$.\cite{Berg}
This issue deserves some further comment.
It has been argued that, in fact, the fundamental concept of quasilocal
energy {\em must} depend on the spanning three-slice, since it is the
spanning slice
which determines the fleet of observers at $B$ (defining an equivalence
class of
spanning three-slices).\cite{BY,BYL,Lauiucaa2} A total energy
contained within the two-surface is associated with each fleet.
Therefore, in the spinorial constructions it seems
that one should use a supplementary spinor equation which does
depend on the spanning three-slice (at least in the sense that on
$B$ one may choose the timelike vector associated with the solution
spinors as a boundary condition). From an alternative viewpoint, if one
employs an ``intrinsic'' supplementary equation which is solved on the
{\em unbounded} two-surface $B$, then one should allow the solution
spinors to {\em select} a preferred equivalence class of spanning slices.

Independent of spinor methods, Brown and York have determined what geometric
entity plays the role of quasilocal energy in general relativity.\cite{BY}
Their
analysis yields an energy surface density which is defined on the bounding
two-surface of a generic spacelike slice. The quasilocal energy, the integral
over the two-surface of the energy surface density, possesses the following
salient properties: (i) it can be derived from a first-principles approach from
the gravitational action via the Hamilton-Jacobi method, (ii) it is the value
of
the on-shell Hamiltonian (defined from the canonical action) corresponding to
the
choice on $B$ of a unit lapse function and zero shift vector, and (iii)
it has been shown to play the role of thermodynamical internal energy in the
context of
black-hole thermodynamics.\cite{BMYW} Furthermore, (iv) the Brown-York
quasilocal energy surface density transforms under switches of the $\Sg$ slice
spanning $B$ (generalized boosts) in a manner which is in full accord with the
equivalence principle.\cite{BYL,Lauiucaa2,Lau} Therefore, it is of interest to
establish
a connection between the Brown-York quasilocal energy and the Witten-Nester
form.
Indeed, such a connection supports the use of the Witten-Nester form (which has
been employed so spectacularly in asymptopia) in the quasilocal context, and,
furthermore, provides one with a non-spinor vantage point from which the
various
supplementary spinor equations and the issues mentioned in the previous
paragraph can
be examined. The supplementary equations studied in some detail here are the
Sen-Witten equation\cite{Witten,Sen,Penrose} and the Dougan-Mason
equation.\cite{Mason}

A key feature of the Brown-York construction (which may be regarded as
property (v)) is that it allows one the freedom to assign the reference point
of
the quasilocal energy. When possible, it is perhaps most natural to reference
the energy against flat space, {\em i.e.} assign flat space the zero value of
energy. This paper establishes the general relationship between the Brown-York
quasilocal energy and the common ``spinorial definition of quasilocal energy''
constructed from the Witten-Nestor two-form. It is shown that spinors may
always be
chosen so that the spinorial definition is the ``unreferenced'' Brown-York
quasilocal energy. However, when the Witten-Nester expression is evaluated on
solution spinors to the Sen-Witten equation (obeying
appropriate boundary conditions), an implicit reference point for the energy is
set. The origin of the reference point is the Sen-Witten equation.
In solving the elliptic
Sen-Witten equation on $\Sigma$, one encounters the seemingly {\em desirable}
type of boundary-value
problem in which the timelike vector on $B$ associated with the solutions
spinors may be
fixed as a boundary condition. However, the solution spinors to the Sen-Witten
equation are altered when the spanning slice $\Sigma$ is perturbed or
``wiggled''
in the ambient spacetime $M$, even if the $\Sigma$ timelike normal $u^{\mu}$ is
held fixed on $B$. Clearly then, an energy expression built with Sen-Witten
spinors
is certainly {\it not} quasilocal in the truest sense, since such an expression
does not depend solely of the fleet of (Eulerian) observers at $B$.\footnote{I
thank
J. Samuel for bringing this point to my attention.}
Nevertheless, as shown later, the Sen-Witten energy expression can always be
written in a form which is essentially identical to the that of the Brown-York
QLE.
Moreover, for the case of round spheres embedded in the preferred time
slices of the
Schwarzschild geometry, the Sen-Witten equation determines the flat-space
reference point. Finally, the general results are also used to further
elucidate
the remarkable features of the (Schwarzschild-case) quasilocal energy of Dougan
and Mason. It is shown that this energy is also equivalent to the
Brown-York quasilocal energy (for an distinguished spanning slice) and that a
flat-space reference, arising from the chosen supplementary equation, is also
implicit in the Dougan-Mason definition.

The organization of this work is as follows. A preliminary section fixes
all of the relevant geometric and spinorial conventions. In addition, this
section collects the basic Brown-York results necessary for the following
discussion. The next section, devoted to spinorial definitions of quasilocal
energy,
examines the general relationship between the Brown-York quasilocal
energy and the Witten-Nester form. This section also applies the general
results to the
specific Schwarzschild case for both the Sen-Witten and Dougan-Mason
equations. For the convenience of the reader the appendix uses the
Newman-Penrose method to present
the general solutions to the Schwarzschild-case Sen-Witten and Dougan-Mason
equations.

\section{Preliminaries}

\subsection{Conventions}
\paragraph{Topology and index conventions}
Consider a generic spatially-bounded spacetime region $M$ which
is topologically the Cartesian product of an orientable three-manifold
$\Sigma$ with spatial boundary $\partial\Sg = B$ and a closed connected
segment of the real line $I$. One need not assume that the boundary $B$ of
the three-space $\Sigma$ is simply connected. The spacetime $M$ is foliated
by a time function $t: M\rightarrow I$. The level hypersurfaces of this
function or {\em slices} are the leaves of the foliation,
and the slices labeled by
the initial and final endpoints of $I$ are denoted respectively by $t'$
and $t''$. The product of $B$ with $I$ is an element of the three-boundary
of $M$ and is denoted by $\cal T$. This element is often referred to as
the three-boundary, even though technically the three-boundary of $M$
consists of $\cal T$, $t'$, and $t''$. One need not consider $\T$ to be a
physical boundary. Rather, one may imagine $\T$ to be just some locus of points
separating $M$ off from an ambient spacetime $\cal U$ (the universe).

Lowercase Greek letters $\mu$, $\nu$, $\sigma, \cdots$ are spacetime $M$
indices.
Lowercase latin letters $i$, $j$, $k, \cdots$ from the middle of the
alphabet are spatial
$\Sigma$ indices, while lowercase latin letters $a$, $b$, $c, \cdots$ from the
first part of the alphabet are spatial $B$ indices.
There is no need to consider $\cal T$ indices.
Orthonormal (or when appropriate pseudo-orthonormal)
labels and indices for each space are represented by the same letters
with hats. For example, $\hmu$ is a spacetime tetrad index and $\ha$
is a $B$ orthonormal index.

\paragraph{Spin conventions}
The $(-,+,+,+)$ metric-signature convention is used throughout this
discussion. Since this is a somewhat uncommon choice when spinor methods
are employed, a few basic conventions are listed now. Using a normalized
spin dyad $\{e_{\bf A} | {\bf A} = 1 , 2\}$ ($e_{1}\,^{A} = o^{A}$,
$e_{2}\,^{A} = \iota^{A}$, and $\av_{AB}\, o^{A}\,\iota^{B} = 1$ where
$\av_{AB}$ denotes skewsymmetric inner product), one can construct the
following spin-tensor fields:

\ba \sigma_{\hat{0}}\,^{AA'} & = & -
\frac{i}{\sqrt{2}}\left(o^{A}\,\overline{o}\,\! ^{A'} +
\iota^{A}\,\overline{\iota}\,\! ^{A'}\right)  \nonumber \\
\sigma_{\hat{1}}\,^{AA'} & = & -
\frac{i}{\sqrt{2}}\left(o^{A}\,\overline{\iota}\,\! ^{A'} +
\iota^{A}\,\overline{o}\,\! ^{A'}\right) \nonumber \\
\sigma_{\hat{2}}\,^{AA'} & = & - \frac{i}{\sqrt{2}}\left(- i\,
o^{A}\,\overline{\iota}\,\! ^{A'} +  i\,\iota^{A}\,\overline{o}\,\!
^{A'}\right) \label{basictensors} \\ \sigma_{\hat{3}}\,^{AA'} & = & -
\frac{i}{\sqrt{2}}\left(o^{A}\,\overline{o}\,\! ^{A'} -
\iota^{A}\,\overline{\iota}\,\! ^{A'}\right)\, .  \nonumber
\ea
The hat on the $\hrho$ index of the $\sigma_{\hrho}\,^{AA'}$ is present because
these indices are to be identified with a spacetime tetrad. The $\hrho$
index is only a label and {\em not} a tensorial index.
Given an orthonormal spacetime cotetrad $\{e^{\hrho}\}$ on $M$ and the set
of $\sigma_{\hrho}\,^{AA'}$ tensors above, one can construct a {\em spacetime
soldering form} (or $SL(2,C)$ {\em soldering form}) on $M$ with the definition

\be e^{AA'}\,_{\mu} := \sigma_{\hrho}\,^{AA'}\, e^{\hrho}\,_{\mu}\, ,
\label{solder}
\ee
Note that the soldering form is imaginary $\bar{e}^{AA'}\,_{\mu} =
- e^{AA'}\,_{\mu}$ and that the following incredibly useful identities
involving the soldering form and its inverse hold:

\ba e^{AA'}\,_{\mu}\, e_{BB'}\,^{\mu} & = &
\av_{B}\,^{A}\,\bv_{B'}\,^{A'} \nonumber \\ e^{AA'}\,_{\mu}\,
e_{AA'}\,^{\nu} & = & g^{\nu}_{\mu}  \label{watermelon} \\
e^{AA'}\,_{\mu}\, e_{BA'}\,_{\nu} & = & \frac{1}{2}\,
g_{\mu\nu}\,\av_{B}\,^{A} - \frac{i}{2}\,\e_{\mu\nu\sigma\rho}\,
e^{AA'}\,^{\sigma}\, e_{BA'}\,^{\rho} \nonumber \\ e^{AA'}\,_{\mu}\,
e_{AB'}\,_{\nu} & = & \frac{1}{2}\, g_{\mu\nu}\,\bv_{B'}\,^{A'} +
\frac{i}{2}\,\e_{\mu\nu\sigma\rho}\, e^{AA'}\,^{\sigma}\,
e_{AB'}\,^{\rho}\, .
\nonumber
\ea
These may be verified by explicit calculation using the full expressions for
the $e_{AA'}\,^{\mu}$ and the formulas
$o_{A}\,\iota^{A} = 1$, $o_{A}\, o^{A} = 0$,
{\em etc.}

\subsection{Brown-York quasilocal energy}

\paragraph{General considerations}
The Brown-York notion of quasilocal energy (often {\em QLE} hereafter)
stems from a Hamilton-Jacobi analysis of a gravitational action functional
suitable for the bounded region $M$. Their analysis in
Ref. \cite{BY} leads to an energy
surface density $\av$, which is defined on $B$ and is associated with
the energy of the $\Sg$ gravitational and matter fields contained within
$B$. The energy surface density has the form

\be
\av = \av_{\it 1} - \av_{\it 0} = \frac{1}{\kappa}\,
\left( k - k_{\it 0}\right)\, , \label{oreo}
\ee
where $\kappa = 8\pi$ (in units with $G = c = 1$) and
$k = \kappa\, \av_{\it 1}$ is the trace $\sigma_{ab}\, k^{ab}$ of the
extrinsic curvature of $B$ as embedded in $\Sg$ which has three-metric
$h_{ij}$. The reference term $k_{\it 0} (\sigma_{ab})$ is an arbitrary quantity
depending on the intrinsic geometry of $B$  (the $B$ metric is $\sigma_{ab}$)
and its presence arises from
the freedom to add boundary terms to the gravitational action which depend
solely on the fixed boundary data. Usually, $k_{\it 0}$ is assumed to be the
flat-space reference, the trace of the extrinsic curvature of a two-surface
which is isometric to the surface $B$ {\em but which is embedded in} $R^{3}$
{\em rather than} $\Sg$. In general such a construction is not possible.
Therefore, for certain two-surfaces it is not possible to define a flat-space
density $\av_{\it 0}$, however, the
unreferenced $\av_{\it 1}$ is always well-defined. The final discussion section
of this work addresses this point more fully. The definition (\ref{oreo})
ensures that the total quasilocal energy

\be
E = \int_{B} {\rm d}^{2}x\, \sqrt{\sigma}\,
\varepsilon \label{grapefruit}
\ee
is a functional on the $\Sg$ gravitational phase space. It proves useful
to make the following natural split of the quasilocal energy: $E = E_{\it 1}
- E_{\it 0}$, where

\be
E_{\it 1} = \int_{B} {\rm d}^{2}x\, \sqrt{\sigma}\,\varepsilon_{\it 1}\, ,
\ee
and is referred to hereafter as the {\em unreferenced} QLE.
The reference term (or subtraction term) is

\be
E_{\it 0}  = \int_{B} {\rm d}^{2}x\,
\sqrt{\sigma}\,\varepsilon_{\it 0}\, . \label{scuppernong}
\ee
This work considers $E_{\it 0}$ to be a suitable reference point for the
QLE whenever $k_{\it 0}$ represents the trace of the extrinsic curvature of a
two
surface which is isometric to $B$ but is embedded in some other
three-geometry (preferably, but not necessarily, $R^{3}$). Of course,
this general definition of the reference term is not fully satisfactory,
because there has been no mention of the initial-value constraints.
Some commentary on the role of the constraints in this definition is found
below.

Finally, the later analysis of the Dougan-Mason QLE demands that one consider
a {\em normal momentum density}.\cite{Lau,BYL} Introduce the $B$ density

\be
(j_{\it 1})_{\vdash} = - \frac{1}{\kappa}\, \, \sigma_{ij}\, K^{ij}\, .
\ee
where $K_{ij}$ represents the extrinsic curvature of $\Sg$ as embedded in $M$.
(The ``${\,}_{\it 1}$'' notation has been employed in anticipation of allowing
the freedom to consider a reference-term contribution to this normal momentum
density, in which case
$j_{\vdash} = (j_{\it 1})_{\vdash} - (j_{\it 0})_{\vdash}$.). In a
Hamilton-Jacobi analysis of the gravitational action, $j_{\vdash}$
arises (apart from a factor of $\sqrt{\sigma}$) as minus the variation
of the classical action with respect to a unit stretch of a ``radial''
lapse function $\al$ associated with a $(1+2)$ decomposition of the
metric on either $t''$ or $t'$. This is similar to the origin of
$\varepsilon$, which is (apart from a factor of $\sqrt{\sigma}$)
minus the variation of the classical action with respect to a unit
stretch of a temporal lapse function $N$ associated with a $(2+1)$
decomposition of the $\cal T$ metric.\cite{BY} Ref. \cite{Lau}
and a forthcoming paper \cite{BYL} devoted to ``boosted''
quasilocal stress-energy-momentum examine how $(j_{\it 1})_{\vdash}$
and $\varepsilon_{\it 1}$ behave under switches of the three-slice spanning
$B$.

\paragraph{The Schwarzschild example}
Following Ref. \cite{BY}, one may specialize the preceding results to
the simple yet illustrative case of the Schwarzschild geometry. Work with
the line element
in standard coordinates \cite{MTW},

\be
{\rm d}s^{2} = - N^{2}\, {\rm d}t^{2} + \al^{2}\,  {\rm d}r^{2}
+ r^{2} \left( {\rm d}\theta^{2} + \sin^{2}\theta\, {\rm d}\phi^{2}\right)\, ,
\label{raisen}\ee
where the lapse function is

\be
N = \al^{-1} = \sqrt{1 - \frac{2{\rm M}}{r}}\, .
\ee
The mass of the hole is (``block-Roman'') $\rm M$.
For a constant time slice, the two-surface $B$ is the round sphere
specified by $r = r_{+} > 2{\rm M}$.
For Schwarzschild the quasilocal energy density is

\be
\varepsilon = \varepsilon_{\it 1}  - \varepsilon_{\it 0}  =
\frac{2}{\kappa}\left. \left(- \frac{1}{\al\, r}
+ \frac{1}{r}\right) \right|_{r = r_{+}}\, .
\ee
Notice that the reference term

\be
E_{\it 0}  = \int_{B}{\rm d}^{2}x\,\sqrt{\sigma}
\left( - \frac{2}{\kappa\, r_{+}}\right)
\label{cherry}
\ee
ensures that the the full quasilocal energy
(\ref{grapefruit}),

\be
E = r_{+}\left[ 1 - \left[1 -
\frac{2{\rm M}}{r_{+}}\right]^{1/2}\right]\, ,
\label{bingcherry}
\ee
is referenced against flat-space. In the
limit $r_{+} \rightarrow \infty$ the quasilocal energy $E$ becomes the
Schwarzschild mass $\rm M$.
The momentum density $(j_{\it 1})_{\vdash}$ vanishes
identically for the preferred $t =$ constant Schwarzschild slices.
Nevertheless, for future purposes it is instructive to
retain $(j_{\it 1})_{\vdash}$
in the formalism.

\section{Spinorial quasilocal energy}
This purpose of this section is twofold: (i) to investigate
the general relationship between the Witten-Nester two-form and the
Brown-York QLE density and (ii) to examine
the general relationship for the specific case of the
Schwarzschild geometry. From the results of this section one can also infer
the relationship between the Witten-Nester form and the Brown-York quasilocal
momentum densities $j_{a}$ of Ref. \cite{BY}. However, though there is some
commentary concerning the relationship with the $j_{a}$,
the focus of the discussion rests on the link with the energy surface density
$\varepsilon$.

The sought-for link is made in the following way.
First, $\varepsilon_{\it 1} \, \sqrt{\sigma}\, {\rm d}^{2}x$ and
$(j_{\it 1})_{\vdash}\, \sqrt{\sigma}\, {\rm d}^{2}x$ are expressed as
Sparling two-forms. Once this is achieved, it is quite
straightforward to derive a
formula which relates the Witten-Nester two-form directly to these expressions.
This formula (number (\ref{tulip}) below) has the following form:

\be
E_{s} = \int_{B}{\rm d}^{2}x\,\sqrt{\sigma}\left(\g\,
\varepsilon_{\it 1} - v\,\g\,(j_{\it 1})_{\vdash}\right)
- {\cal E}\, , \label{kiwi}
\ee
where $E_{s}$ is the ``spinorial definition of quasilocal energy'' built
from the Witten-Nester form and associated with a timelike vector
$\bar{u}^{\mu}$
normal to $B$. At $B$ the vector $\bar{u}^{\mu}$ is related to the
future-pointing $\Sg$ normal $u^{\mu}$
by $\bar{u}^{\mu} = \g\, u^{\mu} + v\,\g\, n^{\mu}$, where $n^{\mu}$ is the
normal of $B$ in $\Sg$ and $v$ is a point-dependent boost velocity which
defines a local relativistic factor $\g = (1 - v^{2})^{-1/2}$.
Also, $\cal E$ is an ``anomalous term.'' Note that $E_{s}$ is spinor-dependent,
in the sense that
$E_{s}$ depends on how $\bar{u}^{\mu}$ is broken
down into constituent spinors. Therefore, the anomalous term $\cal E$ is also
spinor-dependent in the same fashion. It is shown that one can easily set $v =
0$ and ${\cal E} = 0$ by ``hand picking'' the spinors (to be inserted into
$E_{s}$) in an obvious way. However, when the energy expression is built from
spinors which obey the Sen-Witten equation (and natural boundary
conditions), though $v$ again vanishes the anomalous term
$\cal E$ serves as a reference point of the energy. Hence, $E_{s}$ with
Sen-Witten spinors has the same form as the full quasilocal energy
(\ref{grapefruit}).

After this general discussion is completed, the following two central results
are given.
First, for the Schwarzschild geometry the anomalous term $\cal E$ in
the expression (\ref{kiwi}) serves as the correct flat-space
subtraction $E_{\it 0}$ from (\ref{cherry}) when the spinors inserted into
$E_{s}$ obey the Sen-Witten equation. Second, though the analysis is more
subtle, precisely the same type of interpretation can made for the Dougan-Mason
Schwarzschild-case QLE.

\subsection{General considerations}
\paragraph{Witten-Nester two-form}
With the $(-,+,+,+)$ metric-signature convention of this work the
{\em Witten-Nester two-form} is written as

\be
F\left[\bar{\eta}\, ,\, \xi\right] =
- \frac{2}{\kappa}\,\,\bar{\eta}_{A'}\,
{\cal D}\xi_{A} \wedge e^{AA'}\, ,
\ee
where $\cal D$ is the spinorial exterior derivative. Its action on
general spin tensors may be inferred from its action on
spin covectors (``cospinor-valued scalars''): ${\cal D}\xi_{A} =
{\rm d}\xi_{A} - \xi_{B}\, A^{B}\,_{A}$, where $A^{A}\,_{B}\,_{\mu}$
are the (unprimed) spin connection coefficients which specify
the $SL(2,C)$ connection. The $SL(2,C)$ spin connection is compatible
with the soldering form in the following sense.

\be
{\cal D}_{\mu}\, e^{AA'}\,_{\lb} =
e^{AA'}\,_{\sigma}\, \Gamma^{\sigma}\,_{\lb\mu}\, .
\label{blackberry}
\ee
The identity (\ref{blackberry}) allows one to show that
${\cal D}\, e^{AA'} = 0$
which implies that the imaginary part of $F[\bar{\xi}\, ,\, \xi ]$
is a pure divergence, and, hence, the integral
$\int_{B}\, F[\bar{\xi}\, ,\, \xi ]$ is a real quantity.

Most spinorial definitions of QLE employ the integrated Witten-Nester
form in a construction of the following type \cite{Berg}:

\be
P^{\underline{AA}'} = -
\frac{2}{\kappa}\int_{B}\bar{\lb}^{\underline{A}'}_{A'}\,
{\cal D}\lb^{\underline{A}}_{A} \wedge e^{AA'}\, . \label{apple}
\ee
(Often only the real part of $P^{\underline{AA}'}$ is considered. Since the
concern
here is only with the energy, the above formula suffices.) The set
$\{\lb^{\underline{A}}_{A}\, |\, \underline{A} = \underline{1}\, ,\,
\underline{2}\}$
comprises a basis for the solution space corresponding to some
supplementary differential equation
${\cal O}\lb^{\underline{A}}_{A} = 0$ ($\cal O$
is some operator linear in ${\cal D}_{\mu}$).
Various combinations of the $P^{\underline{AA}'}$ are
then offered as the value of the energy or momentum associated
with the gravitational and matter fields contained within $B$. In particular,
the spinorial expression for the energy is

\be
E_{s} = \frac{1}{\sqrt{2}}\,\left( P^{\underline{11}'}
+ P^{\underline{22}'}\right)\, . \label{orange}
\ee
At this point, since the spinors which are to be inserted into
$P^{\underline{AA}'}$ are not known, the $E_{s}$ notion of QLE
is not at all well-defined. Notice that this definition is most
natural when the timelike vector associated with the solution
spinors is of unit magnitude on the two-surface $B$.

\paragraph{Sparling two-forms}
The intimate link between the differential Sparling forms and notions
of gravitational energy is well-known \cite{Madore}.
Refs. \cite{Lau,lau} have established the connection between the unreferenced
Brown-York quasilocal energy-momentum
surface densities and the tetrad versions of
the Sparling forms. It is appropriate to briefly recall this correspondence.
First, introduce a spacetime cotetrad
$e^{\hrho}\,_{\mu}$ and its associated connection coefficients
$\Gamma^{\hrho\hsg}\,_{\mu}[e]$. One need only consider the Sparling two-forms

\be
\sigma_{\hrho} \equiv
- \frac{1}{2}\,\e_{\hrho\hsg\htau\hmu}\, \Gamma^{\hsg\htau}
\wedge e^{\hmu}\, ,
\ee
which are completely determined by the cotetrad.
Clearly, these two-forms do not
transform homogeneously under tetrad transformations.
However, the boundary structure of $B$ as embedded in $\Sg \subset M$ provides
a natural (almost unique) gauge. Assume that the time leg $e_{\bot} \equiv
e_{\hat{0}}$ of the tetrad is tied
to the hypersurface normal $u$, and, further, that the third space leg
$e_{\vdash} \equiv e_{\hat{3}}$
is tied to the normal $n$ of $B$ in $\Sigma$. Such a tetrad is said to be
{\em radial time-gauge} or {\em RT-gauge}. Let $s$
denote the inclusion mapping of the
two-surface $B$ in spacetime, $s: B \rightarrow M$. As demonstrated in
\cite{lau}, one can easily verify that the pullbacks to $B$ of $\sigma_{\perp}$
and $\sigma_{\vdash}$ are the following:

\ba
- \frac{1}{\kappa}\, s^{*}\left(\sigma_{\perp}\right)
& = & \varepsilon_{\it 1} \, \sqrt{\sigma}\, {\rm d}^{2}x \nonumber \\
& & \label{pine} \\
\frac{1}{\kappa}\, s^{*}\left(\sigma_{\vdash}\right)
& = & (j_{\it 1})_{\vdash}\, \sqrt{\sigma}\, {\rm d}^{2}x\, , \nonumber
\ea
Likewise, the pullbacks of $1/\kappa\, \sigma_{\ha}$ ($\ha = \hat{1}\, ,\,
\hat{2}$)
are the orthonormal components of the momentum densities $(j_{\it 1})_{a}$ from
Ref. \cite{BY}.

It simplifies matters somewhat -but seems in no way necessary- to work
instead with the complex Sparling two-forms

\be
\sigma^{(+)}\,_{\hrho} \equiv - \e_{\hrho\hsg\htau\hmu}\,
\Gamma^{(+)\hsg\htau}
\wedge e^{\hmu}\, , \label{avocado}
\ee
which are built from the self-dual connection forms

\be
\Gamma^{(+)\hrho\hsg}\,_{\nu} =
\frac{1}{2}\left( \Gamma^{\hrho\hsg}\,_{\nu}
- \frac{{\rm i}}{2}\,\e^{\hrho\hsg\htau\hmu}\,
\Gamma_{\htau\hmu\,\nu}\right)\, .
\ee
Notably, subject to the RT-gauge choice the situation is just as it
was before in (\ref{pine})

\ba
- \frac{1}{\kappa}\, s^{*}\left(\sigma^{(+)}\,_{\perp}\right)
& = & \varepsilon_{\it 1} \, \sqrt{\sigma}\, {\rm d}^{2}x \nonumber \\
& & \nonumber \\
\frac{1}{\kappa}\, s^{*}\left(\sigma^{(+)}\,_{\vdash}\right)
& = & (j_{\it 1})_{\vdash}\, \sqrt{\sigma}\, {\rm d}^{2}x\, . \nonumber
\ea

One can quickly discern the relationship between the Sparling two-forms
and the expression for $P^{\underline{AA}'}$ given in (\ref{apple}).
Let $e_{\bf A}\,^{A}$ (recall that $o^{A} = e_{1}\,^{A}$ and
$\iota^{A} = e_{2}\,^{A}$) be the normalized spin dyad which
corresponds to the RT-gauge tetrad. One has the following formula
between the RT-gauge self-dual connection coefficients and the
spin connection coefficients with respect to the normalized
dyad \cite{Lau,Penrose}:

\be
\Gamma^{(+)\hrho}\,_{\htau\mu}
= e_{\bf AA'}\,^{\hrho}\,
e^{\bf BA'}\,_{\htau}\, A^{\bf A}\,_{{\bf B}\mu}\, .
\label{blueberry}
\ee
Next, define the vector-field $v^{\underline{AA}'\,\mu} =
- {\rm i}\,\lb^{\underline{A}}_{A}\,
\bar{\lb}^{\underline{A}'}_{A'}\,\, e^{AA'\, \mu}$.
Evidently, $v^{\underline{11}'}$ and $v^{\underline{22}'}$ are real
future-pointing
null vectors, while  $v^{\underline{12}'}$ and its complex conjugate
$v^{\underline{21}'}$
are spacelike complex null vectors. Starting from the expression
$- 1/\kappa\,\, v^{\underline{AA}'\, \hrho}\,\,\sigma^{(+)}\,_{\hrho}$
with (\ref{blueberry})
inserted into (\ref{avocado}), one obtains the formula

\be
- \frac{1}{\kappa}\,\, v^{\underline{AA}'\, \hrho}\,\sigma^{(+)}\,_{\hrho}
= \frac{2}{\kappa}\, \bar{\lb}^{\underline{A}'}_{\bf A'}\,
\lb^{\underline{A}}_{\bf B}\, A^{\bf B}\,_{\bf A} \wedge e^{\bf AA'}\, ,
\ee
where along the way one must appeal to the identities (\ref{watermelon}) and
realized that with respect to a normalized dyad the connection coefficients
$A_{{\bf AB}\mu} = A_{{\bf (AB)}\mu}$ are
symmetric in their spin indices. This leads directly to the result

\be
P^{\underline{AA}'} = - \frac{1}{\kappa}\int_{B} v^{\underline{AA}'\,
\hrho}\,\sigma^{(+)}\,_{\hrho}
- \frac{2}{\kappa}\int_{B}\bar{\lb}^{\underline{A}'}_{\bf A'}\,
{\rm d}\lb^{\underline{A}}_{\bf A} \wedge e^{\bf AA'}\, .
\label{prune}\ee
This is a fundamental formula for the considerations of this paper, because
it establishes a connection between the Witten-Nester two-form and the
Brown-York quasilocal surface densities.

To evaluate $P^{\underline{AA}'}$ on general spinors it helps to cast the
anomalous term in (\ref{prune}) in a more convenient form.
Consider the complex null vector
$m^{a} = 1/\sqrt{2}(e_{\hat{1}}\,^{a} + {\rm i}\, e_{\hat{2}}\,^{a})$
associated with the normalized spin frame and its complex conjugate
$\bar{m}^{a}$
which are tangent to $B$ and normalized so that $m^{a}\,\bar{m}_{a} = 1$.
On $B$ the action of the exterior derivative may be written as
${\rm d}f = m[f]\,\bar{m} + \bar{m}[f]\, m$, where as one-forms
$m = m_{a}\, {\rm d}x^{a}$ and $\bar{m} = \bar{m}_{a}\, {\rm d}x^{a}$.
Now the expression (\ref{prune}) for
$P^{\underline{AA}'}$ may be written as

\be
P^{\underline{AA}'} = - \frac{1}{\kappa}\, v^{\underline{AA}'\, \hrho}
\,\sigma^{(+)}\,_{\hrho}
- \frac{2}{\kappa}\int_{B}{\rm d}^{2}x\,
\sqrt{\sigma}\left(\bar{\lb}^{\underline{A}'}_{1'}\,
m\left[\lb^{\underline{A}}_{2}\right] - \bar{\lb}^{\underline{A}'}_{2'}\,
\bar{m}\left[\lb^{\underline{A}}_{1}\right]\right)\, ,
\ee
where the $B$ volume form can also be expressed as
$\sqrt{\sigma}\,{\rm d}^{2}x = {\rm i}\, m \wedge \bar{m}$.
The identity $m \wedge e^{AA'} =
- o^{A}\,\bar{\iota}^{A'}\, \sqrt{\sigma}\,{\rm d}^{2}x$ and its complex
conjugate help in obtaining this last result.

\paragraph{Form of the spinorial QLE}
Adopt (\ref{orange}) as the spinorial notion of QLE, and,
further, assume that the set $\{\lb^{\underline{1}}_{A}\,
,\, \lb^{\underline{2}}_{A}\}$ is subject to the requirement that the
associated timelike vector field  has the form
$\bar{u}^{\mu} = \g\, u^{\mu} + v\,\g\, n^{\mu}$ on $B$.
Subject to this assumption,
the expression for $E_{s}$ is

\ba
\lefteqn{E_{s} = \int_{B}{\rm d}^{2}x\,\sqrt{\sigma}
\left( \g\, \varepsilon_{\it 1} - v\,\g\, (j_{\it 1})_{\vdash}\right) }
& &  \label{tulip}\\
& & - \frac{\sqrt{2}}{\kappa}\int_{B}{\rm d}^{2}x\,
\sqrt{\sigma}\left(\bar{\lb}^{\underline{1}}_{1'}\,
m\left[\lb^{\underline{1}}_{2}\right] - \bar{\lb}^{\underline{1}}_{2'}\,
\bar{m}\left[\lb^{\underline{1}}_{1}\right] + \bar{\lb}^{\underline{2}}_{1'}\,
m\left[\lb^{\underline{2}}_{2}\right] - \bar{\lb}^{\underline{2}}_{2'}\,
\bar{m}\left[\lb^{\underline{2}}_{1}\right]\right)\, . \nonumber
\ea
The integrad of the first term
can also be expressed as $1/\kappa\, \bar{k}$, where $\bar{k}$ is the
extrinsic curvature of $B$ as embedded in a different spanning slice
$\bar{\Sg}$ defined by $\bar{u}^{\mu}$.\cite{Lau,BYL}
(Actually, $\bar{u}^{\mu}$
defines an equivalence
class of slices, since the extension of $\bar{u}^{\mu}$ off $B$ is not
determined.)
Therefore, the spinorial definitions of QLE based on a
construction of the type considered here can be viewed as essentially the
Brown-York notion of
unreferenced QLE plus a contribution from an anomalous term. All of
ambiguity associated with how the vector $\bar{u}^{\mu}$ is
broken down into constituent spinors resides in this anomalous term.
Note that in the anomalous term the components of the
$\lb^{\underline{A}}_{A}$ are with respect to the spin dyad associated with the
RT-gauge tetrad determined by the embedding of $B$ in $\Sg$. It is easy to
write
the anomalous term in terms of $\lb^{\underline{A}}_{A}$ components with
respect
to the dyad determined by the $\bar{\Sg}$ RT-gauge tetrad (and in fact in terms
of these components $\cal E$ has exactly the same form, since $\bar{u}^{\mu}$
is
normal to $B$). Therefore, the need to consider the $(j_{\it 1})_{\vdash}$ term
can be done away with. However, here the slices are ``chosen first'' and
considered primary so the $(j_{\it 1})_{\vdash}$ term is retained.

\paragraph{Relation between $E_{s}$ and $E_{\it 1} $}
Suppose that rather than being determined by a supplementary
equation, the spinors $\lb^{\underline{A}}_{A}$ are constructed
from the normalized spin dyad (which has been tailored to the
$\Sg$ slicing) in the following trivial way. Take
$\lb^{\underline{1}}_{A} = 1/\sqrt{2}(o_{A} +
e^{{\rm i}\psi}\,\iota_{A})$ and
$\lb^{\underline{2}}_{A} = 1/\sqrt{2}(o_{A} - e^{{\rm i}\psi}\,\iota_{A})$
where $\psi$ is some fixed angle (constant on $B$). The null
vectors associated with these spinors are respectively
$1/\sqrt{2}(u + \cos\psi\, e_{\hat{1}} + \sin\psi\, e_{\hat{2}})$
and $1/\sqrt{2}(u - \cos\psi\, e_{\hat{1}} - \sin\psi\, e_{\hat{2}})$.
Notice that by construction (i) the associated timelike vector is the
hypersurface normal $u^{\mu}$ and that (ii) the components
$\lb^{\underline{1}}_{\bf A}$ and
$\lb^{\underline{2}}_{\bf A}$ of these spinors with respect to the
normalized dyad $\{o^{A}\, ,\, \iota^{A}\}$ are constants on $B$.
Therefore, with this
``hand-picked'' selection, the spinorial QLE (\ref{orange}) is
\be
E_{s} = E_{\it 1} =
\int_{B}{\rm d}^{2}x\,\sqrt{\sigma}\, \varepsilon_{\it 1} \, .
\ee

\paragraph{Sen-Witten spinors}
The expression (\ref{tulip}) takes a particularly interesting form
when evaluated on solution spinors to the {\em Sen-Witten equation},
\be
^{\Sg} {\cal D}^{AA'}\, \lb^{\underline{A}}_{A} = 0\, . \label{grass}
\ee
Here the Sen-Witten derivative is defined by
$^{\Sg} {\cal D}_{AA'} \equiv h^{\mu\nu}\, e_{AA'\, \mu}\, {\cal D}_{\nu}$,
where the spacetime expression for the $\Sigma$ metric is
$h_{\mu\nu} = g_{\mu\nu} + u_{\mu}\, u_{\nu}$. With the operation of
$^{\Sg} {\cal D}_{\mu} = h_{\mu}^{\nu}\, {\cal D}_{\nu}$ restricted to
unprimed (or primed) spinors, $^{\Sg} {\cal D}_{\mu}$ is the
derivative operator specified by the $\Sg$ Sen
connection which has well-known properties. Assume that the boundary
conditions placed on the solution spinors ensure that their associated
timelike vector

\be
t^{\mu} = - \frac{\rm i}{\sqrt{2}}\left( \lb^{\underline{1}}_{A}\,
\bar{\lb}^{\underline{1}}_{A'} +
\lb^{\underline{2}}_{A}\,
\bar{\lb}^{\underline{2}}_{A'} \right) e^{AA'\,\mu}
\ee
is the hypersurface normal $u^{\mu}$ {\em on the two-surface} $B$
(further, take the solution set to comprise a normalized dyad on $B$).
Since on $B$ one has $- {\rm i}\, o_{A}\,\bar{\iota}_{A'}\, t^{AA'} =
- {\rm i}\, \iota_{A}\,\bar{o}_{A'}\, t^{AA'} = 0$, the expression
(\ref{tulip}) can be written as

\ba
\lefteqn{E_{SW} = \int_{B}{\rm d}^{2}x\,\sqrt{\sigma}\,
\varepsilon_{\it 1} } & &  \label{rose}\\
& & - \frac{1}{\sqrt{2}\,\kappa}\int_{B}{\rm d}^{2}x\,
\sqrt{\sigma}\left(\bar{\lb}^{\underline{1}}_{1'}\,
m\left[\lb^{\underline{1}}_{2}\right] -
\lb^{\underline{1}}_{2}\,
m\left[\bar{\lb}^{\underline{1}}_{1'}\right]
+ \bar{\lb}^{\underline{2}}_{1'}\,
m\left[\lb^{\underline{2}}_{2}\right] -
\lb^{\underline{2}}_{2}\,
m\left[\bar{\lb}^{\underline{2}}_{1'}\right] \right)\, \nonumber \\
& & - \frac{1}{\sqrt{2}\, \kappa}\int_{B}{\rm d}^{2}x\, \sqrt{\sigma}\left(
\lb^{\underline{1}}_{1}\,
\bar{m}\left[\bar{\lb}^{\underline{1}}_{2'}\right]
 - \bar{\lb}^{\underline{1}}_{2'}\,
\bar{m}\left[\lb^{\underline{1}}_{1}\right] +
\lb^{\underline{2}}_{1}\,
\bar{m}\left[\bar{\lb}^{\underline{2}}_{2'}\right]
- \bar{\lb}^{\underline{2}}_{2'}\,
\bar{m}\left[\lb^{\underline{2}}_{1}\right]\right)\, \nonumber
\ea
Inserting full form of the Sen-Witten equation (\ref{honey}) into
(\ref{rose}) and appealing to the imposed boundary condition, one finds that
\ba
E_{SW} & = & \int_{B}{\rm d}^{2}x\,\sqrt{\sigma}\,\varepsilon_{\it 1}
\label{sourcherry}\\
& & \hspace{1cm} - \frac{1}{\kappa}\int_{B}{\rm d}^{2}x\,
\sqrt{\sigma}\,\sqrt{2}
\left[\psi^{-1}\left(\mu + \Delta\log \psi\right) +
\psi^{-1}\left(\rho - D\log \psi\right)\right]\, . \nonumber
\ea
The conformal factor $\psi \equiv (- t_{\mu}\, u^{\mu})^{1/2}$ is
unity on $B$.
Also in this relation,  $\rho$ and $\mu$ are spin coefficients and
$D$ and $\Delta$ are respectively the derivative operators in the directions
$- {\rm i} o^{A}\, \bar{o}^{A'}$ and $- {\rm i} \iota^{A}\, \bar{\iota}^{A'}$.
(These are defined in the appendix equations
(\ref{peanut}) and (\ref{butter}) for the Schwarzschild case, but these
formulas are fully general.) Note that since $B$ is a spacelike two-surface,
$\mu$ and $\rho$ are real \cite{Penrose} and that the derivatives $D$ and
$\Delta$ appear only in a combination $D - \Delta = \sqrt{2}\, n$. Further,
$k = \sqrt{2}(\mu + \rho)$. Therefore,

\be
k_{\it 0} = \sqrt{2}
\left[\psi^{-1}\left(\mu + \Delta\log \psi\right) +
\psi^{-1}\left(\rho - D\log \psi\right)\right]\, ,
\ee
is the trace of the extrinsic curvature of the two-surface $B$ as embedded
in the conformally transformed geometry which has three-metric
$\psi^{2}\, h_{ij}$. Notice
that this is an isometric embedding, because the conformal factor is unity on
$B$. Therefore, $k_{\it 0}$ may be interpreted as a reference-point
contribution to the energy, and, hence, (\ref{sourcherry}) has the form of the
full
QLE expression (\ref{grapefruit}). Since $\psi = 1$ on $B$, another expression
for the energy is

\be
E_{SW} =  \frac{2}{\kappa}\int_{B}{\rm d}^{2}x\,\sqrt{\sigma}\,
n\left[\psi\right]\, ,
\ee
which is reminiscent of the conformal expression for the ADM energy.
\cite{York6} Of course, one is really interested in on-shell
expressions for the energy in which case the pair $(h_{ij}\, ,\, K^{ij})$
obey the initial-value constraints. The expression above is most natural
when the conformal geometry $\hat{h}_{ij} = \psi^{2}\, h_{ij}$ can be
augmented by $\hat{K}^{ij}$ such that the pair $(\hat{h}_{ij}\, ,
\, \hat{K}^{ij})$ also obey the initial-value constraints. In this case,
$E_{SW}$ is the energy difference between two instantaneous ``states''
of the gravitational field.

\subsection{The Schwarzschild example}
Assume that $E_{s}$ is to be evaluated for a round sphere $B$ in the
Schwarzschild
geometry determined by $r = r_{+} > 2{\rm M}$. It is convenient
to set

\be
m = \frac{e^{-{\rm i}\phi}}{\sqrt{2}\,
r_{+}}\left(\frac{\partial}{\partial \theta}
+ \frac{{\rm i}}{\sin\theta}\,\frac{\partial}{\partial \phi}\right) =
- \frac{P}{\sqrt{2}\, r_{+}}\, \frac{\partial}{\partial \zeta}\, ,
\label{apricot}\ee
where the stereographic coordinate \cite{Penrose}
$\zeta = e^{{\rm i}\phi}\,\cot\theta/2$ and $P = 1 + \zeta\bar{\zeta}$.
Therefore, for the round sphere $B$ the formula (\ref{tulip}) can be written as

\ba
E_{s} & = &  \int_{B}{\rm d}^{2}x\,\sqrt{\sigma}
\left( \g\, \varepsilon_{\it 1} - v\,\g\, (j_{\it 1})_{\vdash}\right)
\label{hay} \\
& &  + \frac{1}{\kappa}\int_{B} {\rm d}^{2}x\,\sqrt{\sigma}\,
\frac{P}{r_{+}}\left(\bar{\lb}^{\underline{1}}_{1'}\,
\frac{\partial \lb^{\underline{1}}_{2}}{\partial \zeta} -
\bar{\lb}^{\underline{1}}_{2'}\,
\frac{\partial \lb^{\underline{1}}_{1}}{\partial \bar{\zeta}} +
\bar{\lb}^{\underline{2}}_{1'}\,
\frac{\partial \lb^{\underline{2}}_{2}}{\partial \zeta} -
\bar{\lb}^{\underline{2}}_{2'}\,
\frac{\partial \lb^{\underline{2}}_{1}}{\partial \bar{\zeta}}\right)\, .
\nonumber
\ea
The remainder of this section evaluates this expression on both
the Sen-Witten and Dougan-Mason spinors.

\paragraph{Sen-Witten spinors}
Assume that the spinors $\lb^{\underline{1}}_{A}$
and $\lb^{\underline{2}}_{A}$ are solutions to the Sen-Witten equation.
The equation (\ref{grass}) must be solved on the
Schwarzschild background subject to required boundary conditions on $B$:
(i) $\{\lb^{\underline{1}}_{A} , \lb^{\underline{2}}_{A}\}$ comprise a
normalized spin dyad and (ii) that the associated timelike vector field is the
$\Sigma$ normal $u^{\mu}$. {\em Appendix B.2} establishes that the
following two spinors are solutions to the Sen-Witten equation and
obey these boundary conditions:

\ba
\lb^{\underline{1}}_{A} & = &
\frac{X}{X_{+}}\, \frac{{\rm i}}{\sqrt{P}}\left(\bar{\zeta}\, \iota_{A} +
o_{A}\right) \nonumber \\
& &  \label{moss}\\
\lb^{\underline{2}}_{A} & = & \frac{X}{X_{+}}\,
\frac{{\rm i}}{\sqrt{P}}\left( - \iota_{A} +
\zeta\, o_{A}\right)\, , \nonumber
\ea
where $X = \al^{-2} + 2\,\al^{-1} + 1$ and $X_{+} = X(r_{+})$.
One has the desired normalization

\be
\varepsilon^{AB}\,\lb^{\underline{1}}_{A}\,\lb^{\underline{2}}_{B} =
\left(\frac{X}{X_{+}}\right)^{2}\, ,
\ee
and the associated timelike vector is

\be
t^{\mu} = \left(\frac{X}{X_{+}}\right)^{2}\, u^{\mu}  \, .
\ee
Hence, the conformal factor $\psi = X/X_{+}$. At $r = r_{+}$ the general
considerations above establish that

\be
E_{SW} = E_{\it 1}  - {\cal E}\, ,
\ee
where $E_{\it 1}$ is the specific Schwarzschild expression

\be
E_{\it 1}  =
\int_{B} {\rm d}^{2}x\,\sqrt{\sigma}\left(- \frac{2}{\kappa\,\al_{+}
r_{+}}\right)
\ee
with $\al_{+} = \al(r_{+})$. Furthermore, with the
components from (\ref{moss}) inserted into (\ref{hay}),
direct calculation shows that

\be
{\cal E} = E_{\it 0}  =
\int_{B} {\rm d}^{2}x\,\sqrt{\sigma}\left(- \frac{2}{\kappa\, r_{+}}\right)\, ,
\ee
which is the correct flat-space reference term. Notice that the exact
flat-space density $\varepsilon_{\it 0}$ is recovered. One can also verify
directly that $\psi^{2}\, h_{ij}$ has vanishing curvature.
Since any spinors which obey the demanded boundary conditions on
$B$ yield $E_{\it 1}$
as the leading term in the expression for $E_{s}$, it seems that
in the Schwarzschild
context the Sen-Witten equation is responsible for the flat-space
reference of the QLE.

\paragraph{Dougan-Mason spinors}
Now assume that the spinors are solutions to the (holomorphic-case)
{\em Dougan-Mason equation},

\be
\dl\lb^{\underline{A}}_{A} = 0\, , \label{veldt}
\ee
where $\dl \equiv m^{\mu}\, {\cal D}_{\mu}$. The quasilocal energy
$E$ from (\ref{bingcherry})
is associated with the fleet of observers at $B$ who are instantaneously
at rest in the
spanning $\Sg$ slice (their world lines are the integral curves of $u$).
However, notice
that the $\dl$ operator is insensitive to local boosts of the $B$ timelike
normal, and,
therefore, the Dougan-Mason equation does not depend on the $\Sg$ slice
spanning $B$. In
general one would not expect to find a set of solution spinors for
(\ref{veldt}) which
provide the $\Sigma$ normal $u$ as  their associated timelike vector. Rather,
one should,
perhaps, turn the situation around and let (\ref{veldt}) {\em determine} a
preferred
spanning slice $\bar{\Sg}$. In the Schwarzschild context it is possible to
achieve a
satisfactory interpretation of Dougan-Mason energy from this perspective.

{\em Appendix B.3} presents the general solution to (\ref{veldt})
for the case of round spheres in Schwarzschild. One may choose

\ba
\lb^{\underline{1}}_{A} & = &
{\rm i}\left(\frac{\al}{P}\right)^{1/2}\left(- \al^{-1}\,\zeta\, o_{A}
+ \iota_{A}\right) \nonumber \\
& & \label{fungus} \\
\lb^{\underline{2}}_{A} & = &
{\rm i}\left(\frac{\al}{P}\right)^{1/2}\left( \al^{-1}\, o_{A}
+ \bar{\zeta}\, \iota_{A}\right)\,  \nonumber
\ea
as the the set of solution spinors. This set is normalized dyad,

\be
\varepsilon^{AB}\, \lb^{\underline{1}}_{A}\,
\lb^{\underline{2}}_{B} = 1\, ,
\ee
however, the associated timelike vector,

\be
\bar{u}^{\mu} = \left(\frac{1 + \al^{2}}{2\al}\right) u^{\mu}
+ \left(\frac{1 - \al^{2}}{2\al}\right) n^{\mu}\, ,
\ee
is not the hypersurface normal of a $t =$ constant
slice $\Sigma$. However, it has the boost-form

\be
\bar{u}^{\mu} = \g u^{\mu} + v\,\g n^{\mu}\, ,
\ee
where  $v = (1 - \al^{2})\, (1 + \al^{2})^{-1}$. For each value of $r_{+}$
one just lets $\bar{u}^{\mu}$ {\em select} a new spanning three-slice
$\bar{\Sigma}$.

Inserting the Dougan-Mason spinors in the spinorial QLE expression
(\ref{hay}), one finds

\be
E_{DM} = \int_{B} {\rm d}^{2}x\,\sqrt{\sigma}\left( \g_{+}\, \varepsilon_{\it
1}
- v_{+}\,\g_{+}\, (j_{\it 1})_{\vdash}\right) - E_{\it 0} \, , \label{guava}
\ee
where $v_{+} = v(r_{+})$ and $E_{\it 0}$  has exactly the same form as
before in (\ref{cherry}) (as may be verified by direct calculation).
The above expression can also be written as

\be
E_{DM} = \int_{B}{\rm d}^{2}x\,
\sqrt{\sigma}\,\bar{\varepsilon}_{\it 1}  - \bar{E}_{\it 0}\, ,
\ee
where $\bar{\varepsilon}_{\it 1}  = 1/\kappa\, \bar{k}$
with $\bar{k}$ representing the trace of the extrinsic curvature
of $B$ {\em as embedded in the selected spanning slice $\bar{\Sg}$}.
The first term is the Brown-York unreferenced QLE associated the fleet of
$\bar{\Sg}$ observers at $B$. Further, $\bar{E}_{\it 0}$ is the correct
flat-space
reference term. It is the {\em metric} data of $B$ which is crucial when
determining the reference term ($B$ must be isometrically embedded in
$R^{3}$ in order to obtain $\bar{\varepsilon}_{\it 0}  = 1/\kappa\,\,
\bar{k}_{\it 0}$).
Of course the induced metric on $B$ is the same whether $B$ is viewed as
embedded
in $\Sg$ or $\bar{\Sg}$. Hence, the flat-space reference terms
$\bar{E}_{\it 0}$ and $E_{\it 0}$ should agree.
Of course, for Schwarzschild the $(j_{\it 1})_{\vdash}$ term vanishes, and
one can verify that (with the chosen metric-signature convention)
the expression (\ref{guava}) enjoys the remarkable feature
of giving the value of ${\rm M}$ for any value of $r_{+}$.

The Dougan-Mason equation
(\ref{guava}) determines just the ``right'' slice such that
for each $r_{+}$ the associated Brown-York QLE is $\bar{E} = {\rm M}$.
It is of interest to examine these slices in greater detail.\footnote{For a
single isolated sphere the Dougan-Mason spinors actually determine only an
equivalence class of slices. In determining the new foliation of $M$ below, one
is solving the Dougan-Mason equation on each of a whole family of spheres
in spacetime.} As a one-form
the $\bar{\Sg}$ normal is

\be
- \bar{u}_{\mu}\, {\rm d}x^{\mu} =
\frac{\g}{\al}\left({\rm d}t - v\,\al^{2}\, {\rm d}r\right)\, .
\ee
Defining the new coordinate

\be
\bar{t} = t - {\rm M}\,\log\left(\al^{4} - 1\right)\, ,
\ee
one can rewrite this one-form as

\be
- \bar{u}_{\mu}\, {\rm d}x^{\mu} = \bar{N}\, {\rm d}\bar{t}\, ,
\ee
where $\bar{N} = \g\, N = \g/\al = 1/\bar{\al}$. The spatial slices
selected by the Dougan-Mason equation
(\ref{guava}) are level hypersurfaces of $\bar{t}$.
Consider the coordinate transformation
$(t,r,\theta,\phi) \rightarrow (t',r',\theta',\phi')
= (\bar{t},r,\theta,\phi)$, under which
the line element becomes

\be
{\rm d}s^{2} = - \bar{N}^{2}\, {\rm d}\bar{t}^{2}
+ \left[\bar{\al}\, {\rm d}r - v\,\bar{N}\,{\rm d}\bar{t}\right]^{2}
+ r^{2}\left({\rm d}\theta^{2} + \sin^{2}\theta\, {\rm d}\phi^{2}\right)\, .
\ee
Clearly, the induced positive definite metric on $\bar{\Sg}$ is

\be
\bar{h}_{ij}\, {\rm d}x^{i}\, {\rm d}x^{j} =
\bar{\al}^{2}\, {\rm d}r^{2}
+ r^{2}\left({\rm d}\theta^{2} + \sin^{2}\theta\, {\rm d}\phi^{2}\right)\, .
\ee
For this metric it is straightforward to calculate that

\be
\bar{\av}_{\it 1}  = \frac{1}{\kappa}\, \bar{k} =
- \frac{2}{\kappa\,\bar{\al}_{+}\, r_{+}} =
- \g_{+}\,\, \frac{2}{\kappa\,\al_{+}\, r_{+}}\, ,
\ee
which is in accord with the result (\ref{guava}). Note, however, that the
barred slicing is not a ``rest frame'' in the sense that

\be
(\bar{\jmath}_{\it 1})_{\vdash} = v_{+}\,\g_{+}\,
\frac{2}{\kappa\,\al_{+} r_{+}}\, .
\ee
The relation between the momentum integral

\be
(\bar{P}_{\it 1})_{\vdash} \equiv \int_{B}{\rm d}^{2}x\, \sqrt{\sigma}\,
(\bar{\jmath}_{\it 1})_{\vdash}
\ee
and the Dougan-Mason $P^{\underline{AA}'}$ is not immediate, since
none of the spacelike vectors associated with the
solution spinors are $\bar{n}^{\mu}$, the spacelike normal of $B$ in
$\bar{\Sg}$.

\section{Discussion}

This paper has interpreted in a new light the role played by the
Sen-Witten equation in the spinorial expression for energy.
Arguably, the ``job'' of the Sen-Witten equation is to provide a
definite reference point for the energy. It should be stressed
that the energy expression (\ref{orange}) determined by Sen-Witten
spinors is {\em not} offered as a substitute for the definition of
Brown-York QLE given in (\ref{grapefruit}), since the Sen-Witten
energy expression does not depend solely on the fleet of Eulerian
observers
at $B$. However, the Sen-Witten expression for the energy always
has the same form as the Brown-York expression. Further, as long
as $u^{\mu}$ at $B$ is preserved, perturbations of $\Sigma$ only
affect the reference point of the Sen-Witten energy expression.
For the case of round spheres embedded in the preferred
time slices of the Schwarzschild geometry, the reference point
determined by the Sen-Witten equation is flat-space. Clearly, the
Sen-Witten equation does not determine the flat-space reference
for a generic two-surface embedded in a spatial section of an
arbitrary spacetime, because in general
the spatial section is not conformally flat. It is of interest
to determine
under what criteria the Sen-Witten equation does determine the
flat-space reference. Perhaps, by choosing a special spanning
slice in some way, one could use the spinor machinery as an
effective way to solve the embedding problem and compute the
appropriate flat-space $\varepsilon_{\it 0}  = 1/\kappa\, k_{\it 0}$
(when it is possible to do so).

The Schwarzschild-case QLE of Dougan and Mason has also been viewed from
a new perspective. Using the Sen-Witten equation, one ``hand selects'' the
spanning slice for which the energy is to be evaluated. However, at least for
the Schwarzschild-case, the Dougan-Mason equation selects a distinguished
slice $\bar{\Sg}$ which spans the two-sphere $B$ such that the associated
Brown-York
QLE is the Schwarzschild mass $\rm M$. It is known that in general the
Dougan-Mason construction breaks down for ``exceptional''
two-surfaces.\cite{Mason}
Perhaps, this is related to an inability to define a flat-space
(or otherwise) reference point for the energy. The construction of the
flat-space $\varepsilon_{\it 0}  = 1/\kappa\, k_{\it 0}$ previously
described makes sense only when $B$ can be embedded isometrically in $R^{3}$
(and if the embedding is in a certain sense unique). Riemannian manifolds with
two-sphere topology and everywhere positive curvature may be globally immersed
(an immersion allows self-intersection) in $R^{3}$ (the immersion is unique up
to translations and rotations).\cite{BY,Spivak} This may also be
contrasted with the fact that in order for the (holomorphic-case) Dougan-Mason
QLE to be non-negative, one must assume that the spin coefficient $\rho' = -
\mu$, the convergence of the inward null normal to $B$, must be non-negative.
This implies that the two-surface is ``suitably'' convex or, in other words,
has no indentations.\cite{Mason}

\section{Acknowledgments}
I am grateful for continued encouragement and new insights
from J. W. York and J. D. Brown, and for a number of clarifying
discussions with J. Samuel
about solutions to the Sen-Witten equation. I also thank N. Dadhich, S. Koshti,
D. Lynden-Bell, M. Seriu, and S. Sinha fo helpful remarks.

\appendix
\section{Supplementary spinor equations}
This appendix presents the Schwarzschild-case
solutions to the Sen-Witten and Dougan-Mason spinorial equations.
The vehicle for examining these equations is the Newman-Penrose formalism.
\cite{Penrose} Since the Newman-Penrose formalism is usually
employed with the $(+,-,-,-)$ metric-signature convention, it is helpful to
cast
some of the basic results in the $(-,+,+,+)$ convention adopted here.

\subsection{Newman-Penrose formalism}
\paragraph{Null tetrad}
A convenient null tetrad associated with the line
element (\ref{raisen}) is

\ba
e_{1} = k = \frac{1}{\sqrt{2}}\left( \frac{1}{N}\,\frac{\partial}{\partial t}
+ \frac{1}{\al}\, \frac{\partial}{\partial r}\right) &  &
e_{2} = l = \frac{1}{\sqrt{2}}\left( \frac{1}{N}\,\frac{\partial}{\partial t}
- \frac{1}{\al}\, \frac{\partial}{\partial r}\right) \nonumber \\
& & \label{passionfruit}\\
e_{3} = m = \frac{e^{- {\rm i} \phi}}{\sqrt{2}\,
r}\left(\frac{\partial}{\partial \theta}
+ \frac{{\rm i}}{\sin\theta}\,\frac{\partial}{\partial \phi}\right) & &
e_{4} = \bar{m} = \frac{e^{{\rm i} \phi}}{\sqrt{2}\,
r}\left(\frac{\partial}{\partial \theta}
- \frac{{\rm i}}{\sin\theta}\,\frac{\partial}{\partial \phi}\right)\,
,\nonumber
\ea
where the convention is that barred Greek indices $\bar{\mu}$ represent
null-tetrad indices and labels and run over $(1 , 2 , 3 , 4)$.
This frame has been rigged to ensure that

\be
m = - \frac{P}{\sqrt{2}\, r}\, \frac{\partial}{\partial \zeta}\, ,
\ee
as seen before in (\ref{apricot}).

\paragraph{Associated spin coefficients}
Define the following derivative operators:

\be
D \equiv k^{\mu}\,{\cal D}_{\mu}\,\,\, ; \hspace{5mm}
\dl \equiv m^{\mu}\,{\cal D}_{\mu}\,\,\, ; \hspace{5mm}
\bar{\dl} \equiv \bar{m}^{\mu}\,{\cal D}_{\mu}\,\,\, ; \hspace{5mm}
\Delta \equiv l^{\mu}\,{\cal D}_{\mu}\,\, . \label{peanut}
\ee
Recalling that the $SL(2,C)$ spin connection is compatible
with the affine connection in the sense of (\ref{blackberry}),
one can calculate the spin coefficients,

\ba
\al  & = & - \iota_{A}\, \bar{\dl} o^{A} = \frac{1}{2}\left(\Gamma_{124} +
\Gamma_{434}\right) = - \frac{\zeta}{2 \sqrt{2}\, r} \nonumber \\
& & \nonumber \\
\bt & = & - \iota_{A}\, \dl o^{A} = \frac{1}{2} \left(
\Gamma_{123} + \Gamma_{433}\right) = \frac{\bar{\zeta}}{2\sqrt{2}\, r}\nonumber
\\
& & \nonumber \\
\mu & = & - \iota_{A}\, \dl\iota^{A} = \Gamma_{423}
= - \frac{1}{\sqrt{2}\,\al\, r} \nonumber \\
& &  \label{butter}  \\
\rho & = & - o_{A}\, \bar{\dl} o^{A}
=  \Gamma_{134} = - \frac{1}{\sqrt{2}\,\al\, r}  \nonumber \\
& & \nonumber \\
\g & = & - \iota_{A}\, \Delta o^{A} =
\frac{1}{2}\left(\Gamma_{122} + \Gamma_{432}\right) =
\frac{1}{2\sqrt{2}\,\al}\,\, \frac{{\rm d}}{{\rm d} r}\,\log N \nonumber \\
& & \nonumber \\
\varepsilon & = & - \iota_{A}\, D o^{A} =
\frac{1}{2}\left(\Gamma_{121} + \Gamma_{431}\right) =
\frac{1}{2\sqrt{2}\,\al}\,\, \frac{{\rm d}}{{\rm d} r}\,\log N\, ,  \nonumber
\ea
where the affine connection coefficients with respect to the null frame
(\ref{passionfruit}) are given by
$\Gamma_{\bar{\mu}\bar{\sigma}\bar{\rho}} = e_{\bar{\mu}}\,^{\lb}\,
e_{\bar{\rho}}\,^{\kp}\, \nabla_{\kp}\, e_{\bar{\sigma} \lb}$ and
may calculated by any of a variety of methods. For the chosen null
frame, all other spin coefficients vanish. One should beware that on
the left-hand side of these relations $\al$ denotes a spin coefficient,
while on the right-hand side $\al$ denotes the radial lapse function. Also,
here
$\varepsilon$ is a spin coefficient and not the QLE density.

\subsection{Sen-Witten equation}
Transvecting the Sen-Witten equation $^{\Sg} D^{AA'}\,\lb_{A} = 0$ with
$\bar{o}^{A'}$ and $\bar{\iota}^{A'}$, one obtains the set

\ba
\bar{\dl} \lb_{1} - \frac{1}{2}\left( D - \Delta \right) \lb_{2} +
\lb_{1}\left(-\al + \frac{\pi}{2} - \frac{\nu}{2}\right) +
\lb_{2}\left(\rho - \frac{\varepsilon}{2} + \frac{\gamma}{2}\right)   & = & 0
\nonumber \\
& & \label{honey}\\
\dl \lb_{2} + \frac{1}{2} \left(D - \Delta\right) \lb_{1}
+\lb_{1}\left(- \mu - \frac{\varepsilon}{2} + \frac{\gamma}{2}\right)
+ \lb_{2}\left(\bt + \frac{\kappa}{2} - \frac{\tau}{2}\right) & = & 0\, .
\nonumber
\ea
(As mentioned, the coefficients $\pi$, $\nu$, $\kappa$, and $\tau$ vanish
for the Schwarzschild geometry. With the adopted conventions, the general
expressions for these
coefficients in terms of the spin dyad have the same forms as found in
\cite{Penrose}.)
For the case at hand these equations take the following specific form:

\ba
P\, \frac{\partial \lb_{1}}{\partial \bar{\zeta}}
+ \frac{r}{\al}\,\frac{\partial \lb_{2}}{\partial r}
- \frac{\zeta\,\lb_{1}}{2} + \frac{\lb_{2}}{\al} & = & 0
\nonumber \\
& & \\
P\, \frac{\partial \lb_{2}}{\partial \zeta}
- \frac{r}{\al}\,\frac{\partial \lb_{1}}{\partial r}
- \frac{\bar{\zeta}\, \lb_{2}}{2} - \frac{\lb_{1}}{\al} & = & 0\, , \nonumber
\ea
where $\al$ is the radial lapse in these equations. The ansatz
\footnote{This ansatz is most readily suggested if one works in the
Geroch-Held-Penrose compacted spin coefficient formalism \cite{Penrose}
in the manner of Dougan in Ref. \cite{Dougan}.}

\ba
\lb_{1} & = & f(r)\, P^{- 1/2}\left(a - b\,\bar{\zeta}\right) \nonumber \\
& & \\
\lb_{2} & = & g(r)\, P^{- 1/2}\left(a\, \zeta + b \right) \nonumber
\ea
($f\, ,\, g$ are real functions and $a\, ,\, b \in C$) reduces these equations
to the
following system of coupled ordinary differential equations:

\ba
\frac{\dr g}{\dr r} + \frac{g}{r} - \frac {\al\, f}{r} & = & 0 \nonumber \\
& & \\
\frac{\dr f}{\dr
 r} + \frac{f}{r} - \frac {\al\, g}{r} & = & 0\, . \nonumber
\ea
Viewed as an initial-value problem with data specified at $r = 2 {\rm M}$,
the solution to this set of equations is

\be
\left(\begin{array}{c} f \\ \\ g \end{array}\right) = \left( \begin{array}{cc}
\frac{X}{2} + \frac{4\,{\rm M}^{2}}{2\, X\, r^{2}} &
\frac{X}{2} - \frac{4\,{\rm M}^{2}}{2\, X\, r^{2}} \\
 & \\
\frac{X}{2} - \frac{4\,{\rm M}^{2}}{2\, X\, r^{2}} &
\frac{X}{2} + \frac{4\,{\rm M}^{2}}{2\, X\, r^{2}} \end{array}
\right) \left(\begin{array}{c} f_{0} \\  \\ g_{0} \end{array}\right)\, ,
\ee
where $f_{0} = f(2 {\rm M})$, $g_{0} = g(2 {\rm M})$, and
$X = \al^{-2} + 2\,\al^{-1} + 1$. One should notice that

\be
\lim_{r \rightarrow 2 {\rm M}} X = 1\,\,\, ;\,\,\,
\lim_{r \rightarrow \infty} X = 4\, ,
\ee
and that on the interval $[\, 2 {\rm M} , \infty )$ $X$ is monotonically
increasing. Hence, one finds that the general solution to the Sen-Witten
equation
associated with the Schwarzschild geometry may be written as

\ba
\lb_{1} & = & X\, P^{- 1/2} \left( a - b\,\bar{\zeta}
\right)\left(\frac{f_{0} + g_{0}}{2}\right) +
\frac{4 {\rm M}^{2}}{2 X r^{2}}\, P^{- 1/2}
\left( a - b\,\bar{\zeta} \right)\left(\frac{f_{0} - g_{0}}{2}\right) \nonumber
\\
& & \label{daisy} \\
\lb_{2} & = & X\, P^{- 1/2} \left( a\,\zeta + b
\right)\left(\frac{f_{0} + g_{0}}{2}\right) -
\frac{4 {\rm M}^{2}}{2 X r^{2}}\, P^{- 1/2}
\left( a\, \zeta + b\right)\left(\frac{f_{0} - g_{0}}{2}\right) \nonumber
\ea

One seeks two linearly independent solutions
$\{\lb_{A}^{\underline{A}} | \underline{A} = \underline{1}\, ,\,
\underline{2}\}$
which have the appropriate behavior at the finite radius $r_{+} > 2{\rm M}$.
To obtain the desired behavior, first define $X_{+} = X(r_{+})$ and
set $f_{0} = g_{0} = X_{+}^{-1}$. With this choice, from the set (\ref{moss})
$\lb^{\underline{1}}_{A}$ corresponds to the selection $a = 0$ and $b = {\rm
i}$
while $\lb^{\underline{2}}_{A}$ is determined by $a = {\rm i}$ and $b = 0$.

\subsection{Dougan-Mason equation}
Dougan has given a full treatment \cite{Dougan} of the Dougan-Mason
equation for the case of round spheres, and the method followed here
has been inspired by this treatment. Transvection of $\dl\, \lb_{A} = 0$
by each of the members of the spin dyad yields two coupled equations

\ba
m\left[\lb_{1}\right] - \beta\,\lb_{1}
+ \sigma\,\lb_{2} & = & 0 \nonumber \\
& & \label{banana}\\
m\left[\lb_{2}\right] + \beta\,\lb_{2}
- \mu\,\lb_{1} & = & 0\, , \nonumber
\ea
For the present Schwarzschild case the first
equation is

\be
\frac{\partial}{\partial \zeta}\,\log\,\lb_{1} +
\frac{\bar{\zeta}}{2P} = 0\, .
\ee
Direct integration yields that

\be
\lb_{1} = P^{- 1/2}\left(c + d\,\bar{\zeta}\right)\, ,
\ee
where $c(\bar{\zeta}, r)$ and $d(\bar{\zeta}, r)$ are undetermined
functions of $\bar{\zeta}$ and $r$ (or one may hold that $r = r_{+}$).
Next, insert the result for $\lb_{1}$ into the second equation of
(\ref{banana}) to find (again, $\al$ is the radial lapse below)

\be
\frac{\partial\lb_{2}}{\partial\zeta} - \frac{\bar{\zeta}\lb_{2}}{2P}
- \frac{\left(c + d\,\bar{\zeta}\right)}{\al\,P^{3/2}} = 0\, .
\ee
Integration of this equation gives

\be
\lb_{2} = \al^{-1}\, P^{- 1/2}\left( c\,\zeta - d\right)\, .
\ee
Therefore, the general solution to the (Schwarzschild-case) Dougan-Mason
equation (\ref{veldt}) is

\be
\lb_{A} =  \al^{-1}\, P^{- 1/2}\left( c\,\zeta - d\right)\, o_{A} -
P^{- 1/2}\left(c + d\,\bar{\zeta}\right)\, \iota_{A}\, .
\ee

One seeks a set linearly independent solutions which serves as a normalized
spin dyad. From the set (\ref{fungus}) one sees that $\lb^{\underline{1}}_{A}$
corresponds to the choice $c = - {\rm i}\, \sqrt{\al}$ and $d = 0$, while
$\lb^{\underline{2}}_{A}$ corresponds to the choice $d = - {\rm i}\,
\sqrt{\al}$ and $c = 0$.

\end{document}